\documentclass[prl,twocolumn,showpacs,amsmath,amssymb,superscriptaddress]{revtex4}
\usepackage[pdftex]{graphicx}
\usepackage[outdir=./]{epstopdf}
\usepackage{enumerate,ulem}
\newcommand \beq{\begin{eqnarray}}
\newcommand \eeq{\end{eqnarray}}

\newcommand{\bm}[1]{\boldsymbol{#1}}

\newcommand{\bfr}{\bm{r}}
\newcommand{\bfk}{\bm{k}}

\newcommand{\bfq}{\bm{q}}

\newcommand{\bfs}{\bm{s}}
\newcommand{\bfS}{\bm{S}}
\newcommand{\bfM}{\bm{M}}

\newcommand{\Tr}{\mathrm{Tr}}

\newcommand{\jsd}{J_\mathrm{sd}}

\begin{document}

\title{Spin textures and spin-wave excitations in doped Dirac-Weyl semimetals}
\author{Yasufumi Araki}
\affiliation{Institute for Materials Research, Tohoku University, Sendai 980-8577, Japan}
\affiliation{Frontier Research Institute for Interdisciplinary Sciences, Tohoku University, Sendai 980-8578, Japan}
\author{Kentaro Nomura}
\affiliation{Institute for Materials Research, Tohoku University, Sendai 980-8577, Japan}

\begin{abstract}
We study correlations and magnetic textures of localized spins, doped in three-dimensional Dirac semimetals.
An effective field theory for magnetic moments is constructed
by integrating out the fermionic degrees of freedom.
The spin correlation shows a strong anisotropy,
originating from spin-momentum locking of Dirac electrons,
in addition to the conventional Heisenberg-like ferromagnetic correlation.
The anisotropic spin correlation 
allows topologically nontrivial magnetic excitation textures such as a transient hedgehog state,
as well as the ferromagnetic ground state.
The spin-wave dispersion in ferromagnetic Weyl semimetal also becomes anisotropic,
being less dispersed perpendicular to the magnetization.
\end{abstract}

\maketitle

\textit{Introduction.} --
Spin-spin correlation gives rise to magnetism in materials.
In magnetically-doped metals and semiconductors,
the correlation among localized magnetic moments is mediated by the motion of carriers,
known as Ruderman-Kittel-Kasuya-Yosida (RKKY) interaction \cite{Ruderman_Kittel,Kasuya,Yosida},
which can give rise to a ferromagnetic order.
In the presence of strong spin-orbit coupling,
so-called Van Vleck paramagnetism,
in which mixing of the valence and conduction bands contributes to the magnetic susceptibility \cite{VanVleck},
can also play an important role in inducing a magnetic order.
In magnetically-doped topological insulators,
it has been theoretically and experimentally verified that
a ferromagnetic order is spontaneously induced by the Van Vleck mechanism,
showing the quantum anomalous Hall effect \cite{Liu_2009,Yu_2010,Chen_2010,Abanin_2011,Nomura_2011,Checkelsky_2012,Chang_2013}.

One of the candidate materials that may show both of the mechanisms discussed above
is a magnetically-doped Dirac semimetal.
A three-dimensional (3D) Dirac semimetal manifests gapless linear dispersions,
namely the ``Dirac cone structure'',
doubly degenerate with time-reversal and spatial inversion symmetries \cite{Young_2012}.
A Weyl semimetal is
a new type of topologically protected gapless quantum state,
with either time-reversal or spatial inversion symmetries broken in a Dirac semimetal \cite{Wan_2011,Burkov_2011}.
Recent realizations of 3D Dirac and Weyl semimetals
\cite{Liu_2014,Liu_2014_2,Neupane_2014,Borisenko_2014}
have opened a new way
to access various nontrivial transport properties,
such as the anomalous Hall effect \cite{Burkov_2014,Burkov_2014_2,Fujimoto_Kohno_2014},
the chiral magnetic effect \cite{Vilenkin_1980,Kharzeev_McLerran_Warringa_2008,Fukushima_Kharzeev_Warringa_2008,Kharzeev_2014},
the negative magnetoresistance \cite{Son_Spivak_2013,Burkov_2014_3,QLi_2014,CZhang_2015},
the charge-induced spin torque \cite{Nomura_2015},
and the anomalous magnon electrodynamics \cite{CXLiu_2013,Hutasoit_2014},
arising from the chiral anomaly of the Dirac-Weyl Hamiltonian.

In Dirac and Weyl semimetals, it has been proposed that the RKKY interaction becomes anisotropic,
showing Ising- and Dzyaloshinskii--Moriya-like interaction terms \cite{HRChang_2015,Hosseini_2015}.
Since spin-momentum locking in the Dirac-Weyl Hamiltonian correlates
electron spin degrees of freedom to its motion in real space,
it is possible that the correlation among localized magnetic moments can depend on the spatial configuration.
Such an anisotropic correlation may give rise to nontrivial magnetic textures.


In this Letter,
we construct an effective field theory for localized magnetic moments in doped Dirac-Weyl semimetals,
to clarify the properties of local spin correlations.
Based on this effective field theory,
we investigate the consequent magnetic textures and spin-wave dispersion,
paying attention to the global rotational symmetry of the system.
We show that
the spin correlation gains a spatial anisotropy due to spin-momentum locking,
and that the Van Vleck mechanism strongly contributes to the anisotropy.
Localized magnetic moments become strongly correlated parallel to the local magnetization,
while the correlation is rather weak in the perpendicular direction.
We find that the additional correlation allows several types of topologically nontrivial magnetic textures
as excitations from the ferromagnetic ground state,
such as a ``hedgehog'' around a single point or a ``radial vortex'' around an axis.
We also investigate the properties of spin-wave (magnon) excitation in the ferromagnetic phase.
The magnon excitation discussed here is not a conventional Nambu--Goldstone (NG) mode,
since it accompanies the real-space symmetry as well as the spin symmetry due to spin-momentum locking.
Consequently, the spin-wave dispersion becomes anisotropic,
less dispersed in the direction transverse to the magnetization direction.
We see that the anisotropy is strongly enhanced
when the Fermi level is close to the Dirac or Weyl nodes,
which is a characteristic feature of Van Vleck mechanism \cite{Kurebayashi_2014}.

\textit{Model.} --
We consider here a three-dimensional Dirac semimetal,
described by the effective Hamiltonian
\begin{align}
H_0 = \int d^3 \bfr \ \sum_{\lambda=\pm} \psi_\lambda^\dag(\bfr) \left[ \lambda v_F \bm{\sigma} \cdot \hat{\bm{p}} -E_F \right] \psi_\lambda(\bfr), \label{eq:H0}
\end{align}
in the continuum limit,
where the electron spin $\bm{\sigma}$ is locked to the direction of the momentum $\hat{\bm{p}} = -i\bm{\nabla}$
due to strong spin-orbit coupling.
Here $\psi_\lambda = (\psi_{\lambda \uparrow},\psi_{\lambda \downarrow})^T$ and $\psi_\lambda^\dag$ are
annihilation and creation operators of Dirac electrons with chirality $\lambda = \pm$,
$\bm{\sigma}$ is Pauli matrix in the spin space,
$v_F$ is the Fermi velocity, and $E_F$ is the Fermi energy measured from the Dirac points.

Localized magnetic moments are introduced by the spin operator $\bfS(\bfr)$.
Correlation between the Dirac electrons and the localized magnetic moments is given
by the $s$-$d$ exchange interaction,
\begin{align}
H_I = - \jsd \int d^3\bfr \ \bfS(\bfr) \cdot \bfs(\bfr), \label{eq:HI}
\end{align}
where $\jsd$ is the exchange coupling constant,
and $\bfs = \sum_\lambda \psi_\lambda^\dag (\bm{\sigma}/2) \psi_\lambda$ is the spin operator of Dirac electrons.
We should note that the total Hamiltonian $H_\mathrm{tot} = H_0 + H_I$ is
no longer symmetric under a global SU(2) transformation for both itinerant and localized spins.
However,
it is still symmetric under a simultaneous rotation in both the O(3) real space and the SU(2) spin space,
generated by the total angular momentum $\bm{J}_\mathrm{tot} = \int d^3\bfr [ \bm{l} + \bfs + \bfS ]$,
with $\bm{l} = \sum_\lambda \psi_\lambda^\dag [\bfr \times \hat{\bm{p}}] \psi_\lambda$ the orbital angular momentum of Dirac electrons.
Here we call this continuous symmetry ``spin-orbital rotational'' symmetry.
This symmetry is broken
either by a real-space anisotropy, such as an anisotropy of Dirac cone,
or a magnetic anisotropy,
i.e. an easy magnetization axis/plane or (spontaneous) formation of a ferromagnetic order.

\textit{Effective field theory.} --
Let us construct an effective field theory for the localized magnetic moments
by integrating out the fermionic degrees of freedom.
Here we use a classical field $\bfM(\bfr)$ for local magnetization,
in place of the quantum spin operator $\bfS(\bfr)$.
The free energy $F$ for the local magnetization is obtained by path integral formalism as
\begin{align}
e^{-\beta F[\bfM]} = e^{-\beta F_M[\bfM]} \int \mathcal{D}\psi^\dag \mathcal{D}\psi \ e^{- S_\mathrm{tot}[\psi^\dag,\psi;\bfM]},
\end{align}
where $F_M[\bfM] = (1/2\chi_M) \int d^3\bfr |\bfM(\bfr)|^2$ is the bare free energy
that accounts for the magnetic susceptibility $\chi_M$.
The imaginary-time action $S_\mathrm{tot}$ corresponds to the total Hamiltonian $H_\mathrm{tot}$,
with the field variables $(\psi_\lambda^\dag,\psi_\lambda)$ accompanied with arguments $(\bfr,\tau)$.
The imaginary time $\tau$ runs from $0$ to the inverse temperature $\beta$.
The single-particle action $S_0$ can be diagonalized in the Fourier space,
with the Dirac propagator $G_{0 \lambda}(i\omega_n,\bfk) = [(i\omega_n +E_F) - \lambda v_F \bm{\sigma} \cdot \bfk]^{-1}$,
where $\omega_n = (2n+1)\pi/\beta$ is the fermionic Matsubara frequency.

\begin{figure}[tbp]
\begin{tabular}{cc}
\includegraphics[width=5cm]{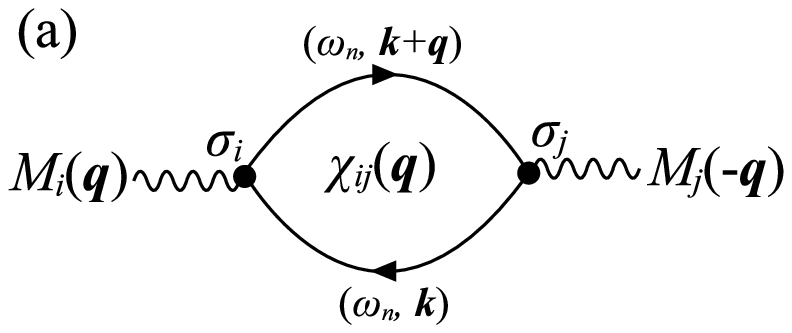}
& \includegraphics[width=2.7cm]{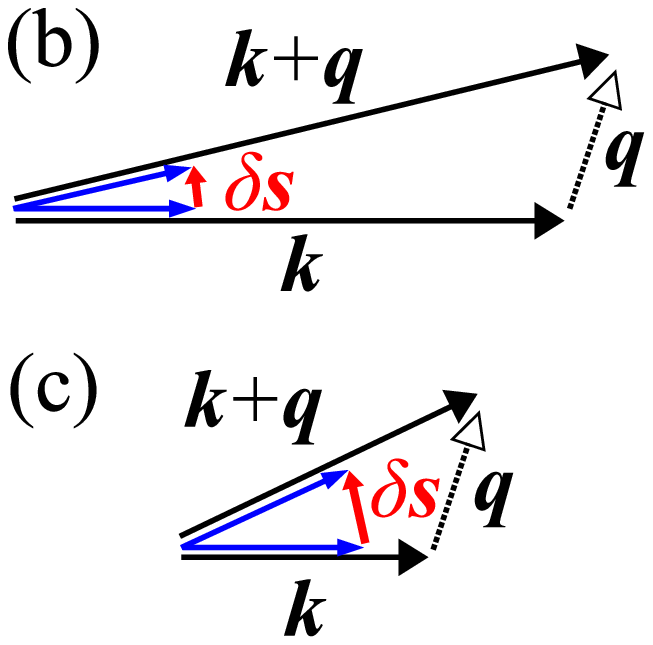}
\end{tabular}
\caption{(a) Feynman diagram of the susceptibility tensor $\chi_{ij}(\bfq)$ defined in Eq.(\ref{eq:susceptibility}).
(b)(c) Schematic pictures of the relationship between the electron momenta $\bfk,\bfk+\bfq$
and the exchanged (spin) angular momentum $\delta \bfs = \bfs_{\bfk+\bfq}-\bfs_{\bfk}$.
$|\delta \bfs|$ becomes relatively larger for smaller $|\bfk| \sim k_F$, as depicted in (c).}
\label{fig:diagrams}
\end{figure}

Taking the local magnetization $\bfM$ perturbatively,
we can evaluate $F[\bfM]$
with the unperturbed Dirac propagator $G_{0\lambda}$ as
\begin{align}
F[\bfM] = F_M[\bfM] - \sum_{\bfq} \sum_{i,j=x,y,z} \frac{\jsd^2}{2} \chi_{ij}(\bfq) M_i(\bfq) M_j(-\bfq). \label{eq:H-eff2}
\end{align}
Contribution from the itinerant electrons starts from $O(M^2)$,
with the ``spin susceptibility'' tensor
 defined by
\begin{align}
\chi_{ij}(\bfq) = \frac{-1}{4\beta V} \sum_{\omega_n,\bfk,\lambda} \Tr\left[\sigma_i G_{0\lambda}(i\omega_n,\bfk+\bfq) \sigma_j G_{0\lambda}(i\omega_n,\bfk) \right], \label{eq:susceptibility}
\end{align}
whose Feynman diagram is given by Fig.\ref{fig:diagrams}(a).
Here $\bfq$ denotes the wavenumber of spin wave in the real space,
and $V$ is the volume of the system.
In order to regularize the integration by $\bfk$,
here we introduce a spherical cutoff with the radius $k_C (\sim 1/a)$,
where $a$ is the characteristic microscopic length scale such as the lattice constant.
Since the quantitative behavior of the ultraviolet (UV) part depends on the shape of the cutoff,
we discuss only the qualitative dependence on $k_C$, with the equivalence sign $\approx$ \cite{Cutoff}.

Due to the spin-orbital rotational symmetry in 3D,
$\chi_{ij}$ can be decomposed into uniform, longitudinal, and transverse parts as
\begin{align}
\chi_{ij}(\bfq) = \chi_0 \delta_{ij} - \Pi_L(q^2) q_i q_j - \Pi_T(q^2) (q^2 \delta_{ij} - q_i q_j). \label{eq:chi-q}
\end{align}
{Using this form,
we can split the free energy $F[\bfM]$ in terms of gradient expansion, with the homogeneous part}
\begin{align}
{F_0[\bfM] = \frac{1}{2}\int d^3\bfr \left(\frac{1}{\chi_M} - \jsd^2 \chi_0 \right) |\bfM(\bfr)|^2}
\end{align}
{and the remaining {gradient} part $F_1[\bfM]$.
The homogeneous part fixes the amplitude of local magnetization $|\bfM|$,
while the gradient part determines the configuration of $\bfM(\bfr)/|\bfM|$, namely the magnetic texture.
Since we are interested in the magnetic texture at a long wavelength,
here we evaluate $F_1[\bfM]$ by power series expansion in $\bfq$.}

Substituting $\chi_{ij}(\bfq)$ up to $O(q^2)$ into Eq.(\ref{eq:H-eff2}),
we obtain the real-space form of the gradient part,
\begin{align}
F_1[\bfM] = \int d^3\bfr \left( J_\mathrm{i} \left[\bm{\nabla} \bfM\right]^2 +J_\mathrm{a} \left[\bm{\nabla}\times \bfM\right]^2 \right). \label{eq:H-eff}
\end{align}
This form
consists of the isotropic term proportional to the square of gradient,
$ [\bm{\nabla}\bfM]^2 = \sum_{ij}(\partial_i M_j)^2$, with the coefficient $J_\mathrm{i}= \jsd^2 \Pi_L(0)/2$,
and the anisotropic term depending on the square of vorticity,
$ [\bm{\nabla}\times\bfM]^2 = \sum_i (\bm{\nabla}\times\bfM)_i^2$, with the coefficient $J_\mathrm{a}=\jsd^2[\Pi_T(0)-\Pi_L(0)]/2$.
The coefficients are given at temperature $T=0$ \cite{Suppl} as
\begin{align}
{J_\mathrm{i} \approx \frac{\jsd^2}{48 \pi^2 v_F^3}\frac{1}{10}, \quad
J_\mathrm{a} \approx \frac{\jsd^2}{48 \pi^2 v_F^3}\left(\ln\frac{k_C}{k_F} -\frac{11}{15}\right).}
\label{eq:J-eff}
\end{align}
The logarithmic factor
\begin{align}
\ln \frac{k_C}{k_F} = \ln \frac{v_F k_C}{E_F} \nonumber
\end{align}
comes from the integration over all the occupied states in the valence and conduction bands.
Such an interband contribution is generated by Van Vleck mechanism,
in which the interband matrix element
contributes to the magnetic susceptibility  due to the strong spin-orbit coupling.
On the other hand, the non-logarithmic part in both $J_\mathrm{i}$ and $J_\mathrm{a}$
comes from the carriers at the Fermi surface.
Such an intraband contribution corresponds to Pauli paramagnetism,
namely the RKKY interaction mediated by carriers.

The dominance of the interband (Van Vleck) contribution depends on the Fermi level $E_F$ \cite{Kurebayashi_2014}.
It becomes stronger at a smaller Fermi momentum,
due to the large angular momentum exchange:
{the angular momentum exchanged between two magnetic impurities via an electron} is equal to the difference $\delta \bfs$
between the incoming electron spin $\bfs_{\bfk} = \lambda \bfk/|\bfk|$ and the outgoing one $\bfs_{\bfk+\bfq} = \lambda (\bfk+\bfq)/|\bfk+\bfq|$
{in the scattering process at a magnetic impurity},
which becomes larger when $|\bfk|$ is small (see Fig.\ref{fig:diagrams}(b)(c)).
This argument does not apply if the Fermi level is extremely close to the Dirac nodes, i.e. $|\bfq| \gtrsim k_F$;
here the $\bfq$-expansion is not applicable due to the non-analyticity at $\bfq=0$,
and $\ln k_F^{-1}$ in the interband contribution gets replaced by $\ln q^{-1}$ \cite{Suppl}.


\textit{Magnetic textures.} --
What does the structure of this effective field theory imply?
The isotropic term in Eq.(\ref{eq:H-eff}) tends to suppress the gradient $\bm{\nabla}\bfM$,
leading to a Heisenberg-like uniform ferromagnetic order,
which comes only from the intraband part.
The anisotropic term, on the other hand, tends to suppress the vorticity $\bm{\nabla}\times \bfM$
as long as $J_\mathrm{a}>0$,
which comes from both the interband and intraband contributions.
The total free energy $F[\bfM]$ prefers the ferromagnetic ground state
at zero temperature.

\begin{figure}[tbp]
\includegraphics[bb=0 0 811 438,width=7cm]{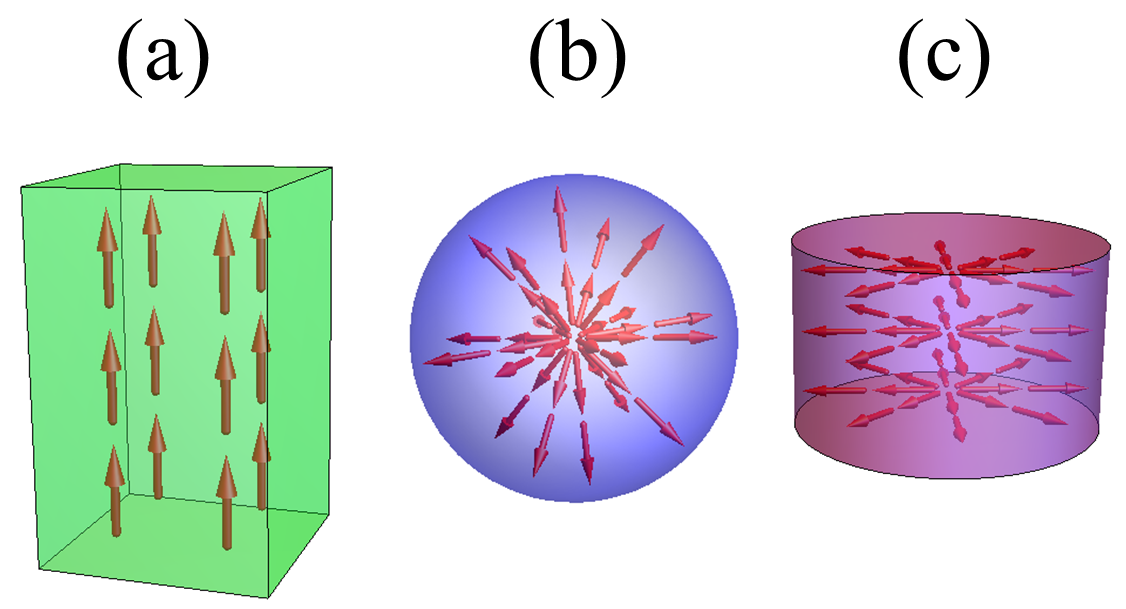}
\caption{Schematic pictures of possible magnetic textures satisfied by the effective field theory:
(a)uniform, (b)hedgehog, and (c)radial vortex structures.
{All of them are degenerate under the interband contribution,
while the intraband contribution splits the degeneracy and chooses (a) as the ground state.}
}
\label{fig:magnetization}
\end{figure}

If the interband contribution is dominant over the intraband contribution,
i.e. the Fermi level is close to the Dirac nodes,
the anisotropic term restricts the magnetic textures of possible excited states.
This term favors vortex-free configurations,
where the local magnetization can be written by using a certain ``scalar potential'' $\phi_m(\bfr)$ as $\bfM = \bm{\nabla}\phi_m$.
It strongly requires the local magnetic moments to be aligned along a line,
while it does not correlate the magnetic moments located perpendicular to the local magnetization,
due to spin-momentum locking.
Such kind of correlation can be regarded as ``Ising-type'' \cite{HRChang_2015,Hosseini_2015},
in the sense that the quantization axis is taken parallel to the relative coordinate between two magnetic moments.
There can be many types of configurations that satisfy this requirement;
the uniform order $\bfM(\bfr)=\bfM_0$ is the simplest case,
while we can take a ``hedgehog'' configuration around a point $\bm{R}$, given by $\bfM(\bfr) \propto (\bfr-\bm{R})/|\bfr-\bm{R}|$,
or a ``radial vortex'' around an axis $(x,y)=(X,Y)$, given by $\bfM(\bfr) \propto (x-X,y-Y,0)/\sqrt{(x-X)^2+(y-Y)^2} $ (see Fig.~\ref{fig:magnetization}).

The intraband contribution gives rise to the isotropic correlation,
which requires a finite excitation energy for the topologically nontrivial textures discussed above,
compared to the ferromagnetic ground state.
As for a hedgehog structure, for instance,
the energy cost per one hedgehog is $\delta E \approx 8\pi J_\mathrm{i} M^2 (L-a)$,
while it gains the entropy $\delta S \approx \ln (L^3/a^3)$.
Hence the entropy contribution to the free energy $\delta F = \delta E - T \delta S$
cannot overcome the energy cost as long as $J_\mathrm{i}>0$ for $L\rightarrow \infty$.
In a finite-size system, on the other hand,
we can expect a crossover at finite temperature,
between the ferromagnetic ground state and the excited state with topological defects satisfying $\boldsymbol{\nabla} \times \bfM =0$.


\textit{Spin-wave excitations.} --
{Let us go back to the ferromagnetic ground state at zero temperature.}
In conventional ferromagnets,
spin SU(2) symmetry is spontaneously broken by the order,
and a spin-wave mode arises as a fluctuation around the mean field.
In Dirac semimetals, on the other hand,
the order breaks the spin-orbital rotational symmetry,
which accompanies the real-space symmetry as well as the spin space.
Thus the spin-wave mode in this system is conceptually different from conventional ones.
Here we search for any consequence of this difference.

Provided that the local spins are fully polarized in the $z$-direction,
the magnetization $\bfM = M\hat{z}$ serves as a mean field through the exchange coupling,
with the splitting energy $\Delta = \jsd M$.
The quantum dynamics of the magnon modes is described
by the bosonic field operators $(z^\dag,z)$,
introduced by the Holstein--Primakoff transformation
$S_z(\bfr) = M - z^\dag(\bfr)z(\bfr), \ S_+(\bfr) \simeq \sqrt{2M}z(\bfr),\ S_-(\bfr) \simeq \sqrt{2M}z^\dag(\bfr)$.
Here the approximation with $\simeq$ is justified
{as long as the fluctuation $z^\dag z$ is sufficiently small}.

The effective field theory for the magnon modes can again be obtained,
by integrating out the fermionic degrees of freedom in the path integral formalism,
as done in Ref.~\cite{Konig_Jungwirth_MacDonald_2001} for dilute magnetic semiconductors.
Here we embed the exchange coupling between the fermions and the mean-field magnetization,
given by $-(\Delta/2) \sum_\lambda \psi_\lambda^\dag \sigma_z \psi_\lambda$,
into the single-particle Hamiltonian,
leaving the spin wave fluctuation part as a perturbation.
Extracting the contribution up to the bilinear in the fluctuation field $z^{(\dag)}$,
we obtain an effective action for the magnons,
\begin{align}
& S_\mathrm{eff}[z^\dag,z] \! = \! \frac{1}{2} \! \sum_{\nu_m,\bfq} \! Z^\dag(i\nu_m,\bfq) {D}_M^{-1}(i\nu_m,\bfq) Z(i\nu_m,\bfq)  \\
& {D}_M^{-1}(i\nu_m,\bfq) = \nonumber \\ 
& \ \left(
\begin{array}{cc}
-i\nu_m + \mu_z q_z^2 + \mu_\perp q_\perp^2 & \mu'_\perp q_-^2 \\
\mu'_\perp q_+^2 & i\nu_m + \mu_z q_z^2 + \mu_\perp q_\perp^2
\end{array}
\right),
\end{align}
where $q_\pm = q_x \pm i q_y$, $q_\perp = |q_\pm|$,
and $Z(i\nu_m,\bfq) = [z(i\nu_m,\bfq),z^\dag(-i\nu_m,-\bfq)]^T$.
The dispersion relation is given by the pole of the propagator $D_M(\epsilon,\bfq)$ as
\begin{align}
\epsilon(\bfq) = \left[(\mu_z q_z^2 + \mu_\perp q_\perp^2)^2 - (\mu'_\perp q_\perp^2)^2 \right]^{1/2} +O(q^4).
\end{align}
Taking the exchange splitting $\Delta$ perturbatively,
the effective Hamiltonian can be constructed from the susceptibility tensor $\chi_{ij}(\bfq)$ defined in Eq.(\ref{eq:susceptibility}).
Limiting the magnon momentum $\bfq$ smaller enough than the Fermi momentum $k_F$,
the coefficients can be evaluated as
\begin{align}
\mu_z = M(J_\mathrm{a} + J_\mathrm{i}), \quad \mu_\perp = \frac{M}{2}(J_\mathrm{a} + 2J_\mathrm{i}),
\quad \mu'_\perp = \frac{M}{2} J_\mathrm{a}, \nonumber
\end{align}
using the effective exchange coefficients $J_\mathrm{i}$ and $J_\mathrm{a}$ in Eq.(\ref{eq:J-eff})
\cite{Suppl}.

The magnon dispersion obtained here is gapless, i.e. $\epsilon(\bfq=0)=0$.
Although the magnon mode discussed here is not
a traditionally-known Nambu--Goldstone (NG) mode
related to spontaneous breaking of internal (e.g.~spin) symmetries,
as discussed above,
it still keeps the gapless structure.
The emergence of such unconventional NG modes, related to real-space symmetries, has recently been discussed
in several literatures \cite{Watanabe_Brauner_2012,Kobayashi_Nitta_2014}.
An intrinsic real-space anisotropy,
such as the Fermi velocity anisotropy or the cutoff anisotropy,
can open a gap in the magnon spectrum,
which possibly occurs in realistic crystalline systems.

\begin{figure}[tbp]
\begin{tabular}{ccccc}
(a) & & (b) & & (c) \\
\includegraphics[bb=0 0 900 900,width=2.5cm]{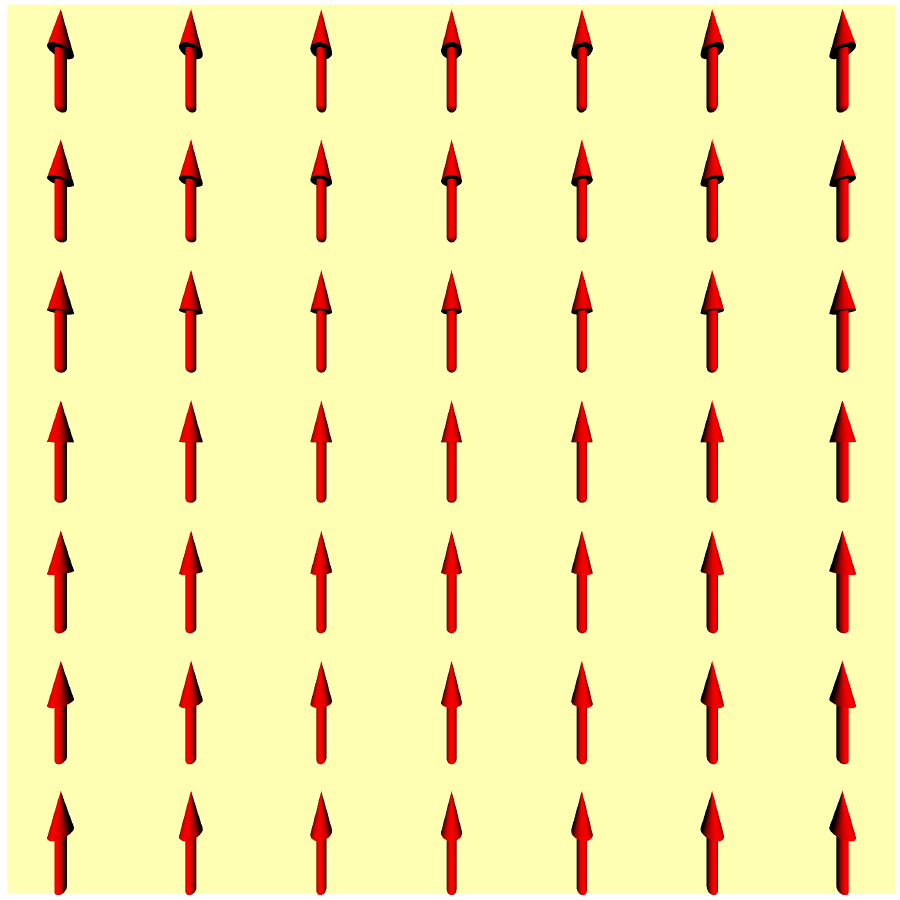} & &
\includegraphics[bb=0 0 900 900,width=2.5cm]{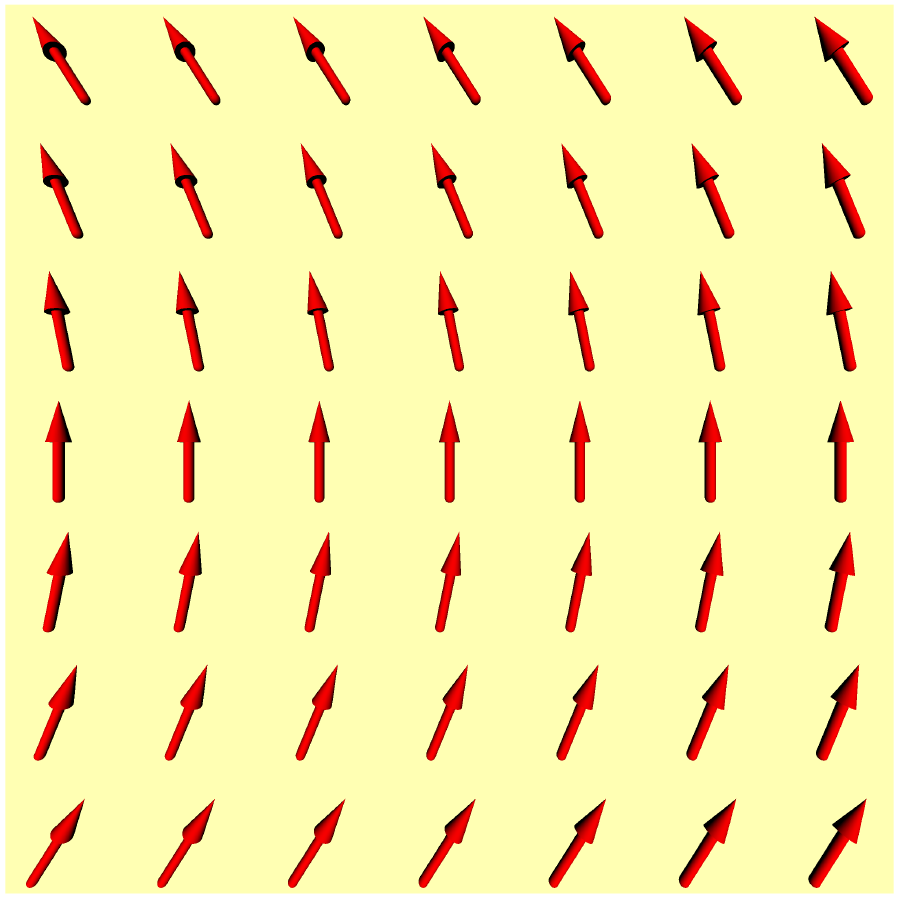} & &
\includegraphics[bb=0 0 900 900,width=2.5cm]{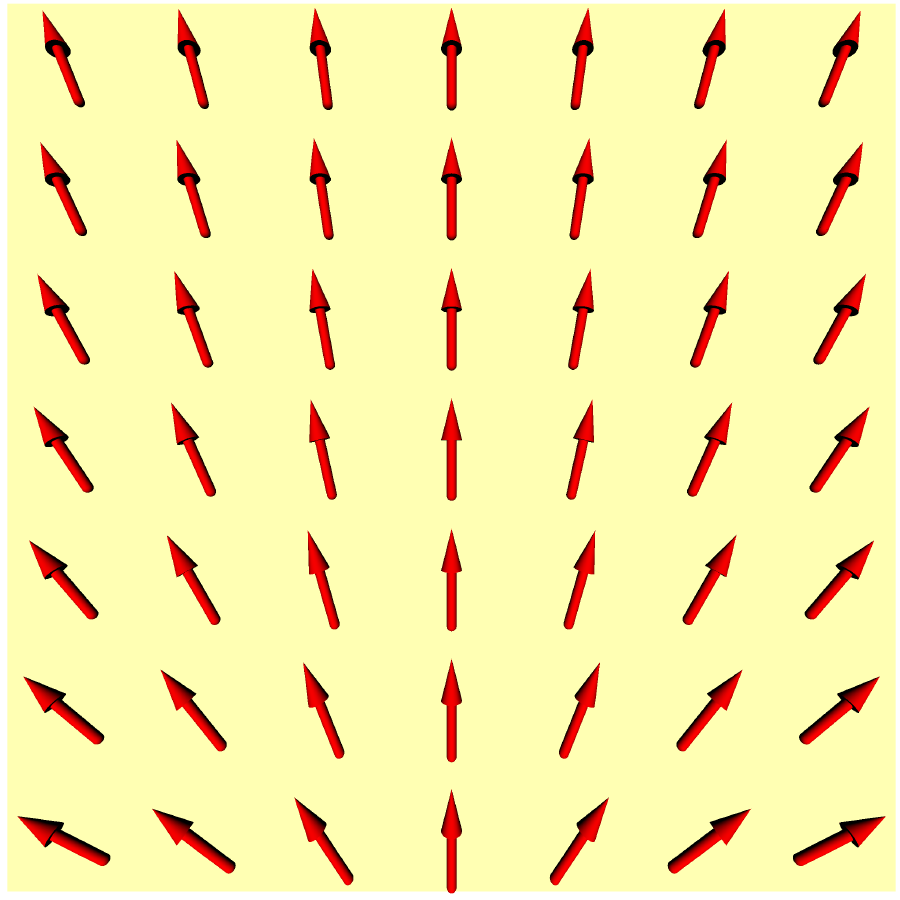}
\end{tabular}
\caption{Schematic pictures of the spin wave excitations,
with red arrows representing the local magnetization.
(a) The uniform ferromagnetic phase.
(b) Longitudinal spin-wave configuration, which is against the longitudinal spin rigidity required by the effective Hamiltonian.
(c) Transverse spin-wave configuration, which does not violate the longitudinal spin rigidity.}
\label{fig:spinwave}
\end{figure}

The consequence of spin-momentum locking appears in the spin-wave dispersion,
starting from the quadratic order in $q$.
The dispersion of longitudinally-propagating mode becomes $\epsilon(q_z) = \mu_z q_z^2$,
while the transverse one reads $\epsilon(q_\perp) =  \tilde{\mu}_\perp q_\perp^2$ with $\tilde{\mu}_\perp =\sqrt{\mu_\perp^2 - \mu'^2_\perp}$.
If the interband (Van Vleck) contribution to the magnetism is larger than the Pauli contribution,
i.e. if $J_\mathrm{a} \gg J_\mathrm{i}$,
there arises a strong anisotropy,
\begin{align}
\mu_z = M(J_\mathrm{a} + J_\mathrm{i}) \gg \tilde{\mu}_\perp = M\sqrt{J_\mathrm{i}(J_\mathrm{a}+ J_\mathrm{i})}.
\end{align}
Here the spin-wave dispersion becomes almost flat in the transverse direction,
i.e. $\epsilon(q_\perp) \sim 0$.
In other words,
as long as the longitudinal correlation is kept rigid {by $\boldsymbol{\nabla} \times \bfM =0$},
the energy cost to modulate the magnetization in the transverse direction is relatively small,
 as shown schematically in Fig.~\ref{fig:spinwave}(c),
while the longitudinal spin wave violates the longitudinal correlation, as shown in Fig.~\ref{fig:spinwave}(b).
This is the straightforward consequence from the effective field theory obtained in Eq.(\ref{eq:H-eff}),
since all the longitudinally correlated configurations are almost degenerate for $J_\mathrm{i} \ll J_\mathrm{a}$.

\textit{Summary and Discussions.} --
We have studied the characteristics of correlation
among localized magnetic moments in Dirac and Weyl semimetals,
{in terms of} the effective field theory {obtained by} integrating out the fermion fields.
Spin-momentum locking leads to various unconventional magnetic properties,
namely the strong longitudinal spin correlation, possibility of topologically nontrivial excitations,
 and the anisotropic dispersion of spin waves in the ferromagnetic phase.
The emergence of such unconventional effects depends on the magnitude of interband (Van Vleck) contribution
characterized by the logarithmic factor $\kappa \equiv \ln(k_C/k_F)$,
which can be tuned through carrier doping.
If the Fermi level is close to the Dirac point,
the interband contribution becomes dominant,
which makes the correlation anisotropies found here quite strong.

Tuning of the chemical potential in Dirac semimetals is realized by introducing dopants;
in Cd$_3$As$_2$ observed in Ref.~\cite{Liu_2014_2}, for instance,
$E_F$ is successfully tuned by alkaline metal doping such as K, up to $E_F^\mathrm{max} = 0.25\mathrm{eV}$ beyond the Dirac point.
Using the in-plane lattice constant $a=4.6\mathrm{\AA}$ and the Fermi velocity $v_F = 1.3 \times 10^6 \mathrm{m/s}$,
the lower bound of $\kappa$ can be estimated as
$\kappa_\mathrm{min} = \ln(v_F a^{-1}/E_F^\mathrm{max}) \sim 2.0$,
where $J_\mathrm{i}$ and $J_\mathrm{a}$ become quite comparable.
The upper bound $\kappa_\mathrm{max}$,
which becomes infinite in the ideal case $k_F=0$,
can be reduced by {the typical long-range length scales},
such as the size of the sample, temperature, the spin coherence length under disorder, etc.
It can be suppressed by the intrinsic breakdown of the spin-orbital rotational symmetry as well,
due to the lattice structure of the system.
We still expect that the upper bound $\kappa_\mathrm{max}$ is quite larger than the lower bound $\kappa_\mathrm{min}$,
which may enable us to observe the crossover between the interband-dominant regime and the trivial regime
by carrier doping.



The authors are thankful to Joseph Barker, Takahiro Chiba, Yusuke Nishida, and Akihiko Sekine for fruitful discussions and comments.
This work was supported by Grant-in-Aid for Scientific Research
 (Grants No.~15H05854, No.~26107505, and No.~26400308) 
from the Ministry of Education, Culture, Sports, Science and 
Technology (MEXT), Japan.


\begin{widetext}

\newpage


\setcounter{figure}{0}   \renewcommand{\thefigure}{S\arabic{figure}}

\setcounter{equation}{0} \renewcommand{\theequation}{S.\arabic{equation}}


\begin{center}
\textbf{\large Supplemental Material}
\end{center}

In this Supplemental Material,
we show some detailed calculation of the effective field theory for background magnetization and the spin-wave dispersion.

\section{Effective field theory}
The effective action for the background magnetization can be obtained by evaluating the spin susceptibility tensor
\begin{align}
\chi_{ij}(\bfq) = -\frac{1}{4\beta V} \sum_{\lambda,\omega_n,\bfk} \Tr \left[\sigma_i G_{0\lambda}(i\omega_n,\bfk+\bfq/2) \sigma_j G_{0\lambda}(i\omega_n,\bfk-\bfq/2) \right],
\end{align}
where we have shifted the momentum $\bfk$ by $\bfq/2$ from the definition in Eq.(\ref{eq:susceptibility}).
Here $G_{0\lambda}$ is the single-particle Green's function (propagator) of the Dirac electron with chirality $\lambda$,
with the matrix structure
\begin{align}
G_{0 \lambda}(i\omega_n,\bfk) &= D_0(i\omega_n,\bfk) \left[ (i\omega_n+E_F) + \lambda v_F \bfk \cdot \bm{\sigma} \right], \\
D_0(i\omega_n,\bfk) &= \left[(i\omega_n+E_F)^2 - |v_F \bfk|^2\right]^{-1}.
\end{align}
Separating the denominator and the numerator of the Green's function,
we can evaluate the trace in the susceptibility tensor,
\begin{align}
\chi_{ij}(\bfq) &= -\frac{1}{4\beta V} \sum_{\lambda,\omega_n,\bfk} D_0(i\omega_n,\bfk+\tfrac{\bfq}{2}) D_0(i\omega_n,\bfk-\tfrac{\bfq}{2}) \nonumber \\
 & \quad \quad \quad \quad \times \Tr \left( \sigma_i \left[(i\omega_n+E_F)+\lambda v_F (\bfk+\tfrac{\bfq}{2})\cdot {\bm{\sigma}}\right] \sigma_j \left[(i\omega_n+E_F)+\lambda v_F (\bfk-\tfrac{\bfq}{2})\cdot {\bm{\sigma}}\right] \right) \\
 &= -\frac{1}{2\beta V} \sum_{\lambda,\omega_n,\bfk} \left( \left[(i\omega_n+E_F)^2 - v_F^2(k^2+\tfrac{1}{4}q^2) \right]^2 - [v_F^2 \bfk \cdot \bfq]^2 \right)^{-1} \nonumber \\
 &\quad \quad \times \left[ (i\omega_n+E_F)^2 \delta_{ij} -(i\omega_n+E_F)\lambda v_F q_k i\epsilon_{ijk} +2v_F^2(k_i k_j - \tfrac{1}{4}q_i q_j) -v_F^2(k^2-\tfrac{1}{4}q^2)\delta_{ij} \right] \\
 &= -\frac{1}{\beta V} \sum_{\omega_n,\bfk} \frac{(i\omega_n+E_F)^2 \delta_{ij} +2v_F^2(k_i k_j - \tfrac{1}{4}q_i q_j) -v_F^2(k^2-\tfrac{1}{4}q^2)\delta_{ij}}{\left[(i\omega_n+E_F)^2 - v_F^2(k^2+\tfrac{1}{4}q^2) \right]^2 - [v_F^2 \bfk \cdot \bfq]^2}
\end{align}
where we have used the relations for Pauli matrices,
\begin{align}
\Tr[\sigma_i \sigma_j] = 2\delta_{ij}, \quad \Tr[\sigma_i \sigma_j \sigma_k] = 2i\epsilon_{ijk}, \quad
\Tr[\sigma_i \sigma_j \sigma_k \sigma_l] = 2(\delta_{ij}\delta_{kl}+\delta_{il}\delta_{jk}-\delta_{ik}\delta_{jl}).
\end{align}

Here we limit ourselves to the case $|\bfq|\ll k_F$,
and evaluate the susceptibility tensor by power series expansion by $\bfq$ up to $O(q^2)$.
Since the $q$-expansion of the denominator reads
\begin{align}
D_0(i\omega_n,\bfk+\tfrac{\bfq}{2}) D_0(i\omega_n,\bfk-\tfrac{\bfq}{2}) &= \left( \left[ D_0^{-1}(i\omega_n,\bfk) - \tfrac{1}{4}v_F^2 q^2 \right]^2 - [v_F^2 \bfk\cdot\bfq]^2 \right)^{-1} \\
 &= D_0^2(i\omega_n,\bfk) + \frac{1}{2} v_F^2 q^2 D_0^3(i\omega_n,\bfk) + [v_F^2 \bfk\cdot\bfq]^2 D_0^4(i\omega_n,\bfk) + O(q^4),
\end{align}
the $q$-expansion of the susceptibility tensor up to $O(q^2)$ is given by
\begin{align}
\chi_{ij}(\bfq) &= \chi_{ij}^{(0)} + \chi_{ij}^{(2)}(\bfq) + O(q^4) \\
\chi_{ij}^{(0)} &= -\frac{1}{\beta V} \sum_{\omega_n,\bfk} D_0^2(i\omega_n,\bfk) \left[ (i\omega_n+E_F)^2 \delta_{ij} +v_F^2(2k_i k_j -k^2\delta_{ij}) \right] = -\frac{1}{\beta V} \sum_{\omega_n,\bfk} \left[ D_0 + D_0^2 2v_F^2 k_i k_j \right] \\
\chi_{ij}^{(2)}(\bfq) &= -\frac{1}{\beta V} \sum_{\omega_n,\bfk} \left[ D_0^2 v_F^2 \left(\frac{3}{4} q^2 \delta_{ij} - \frac{1}{2} q_i q_j \right) + \frac{2}{3} D_0^3 (v_F k)^2 v_F^2 q^2 \delta_{ij} + \frac{2}{15}D_0^4 (v_F k)^4 v_F^2(q^2 \delta_{ij} + 2q_i q_j) \right],
\end{align}
where we have averaged the angular part by the spherical symmetry in the denominator.
Thus $O(q^2)$ term can be decomposed into the longitudinal and transverse terms,
\begin{align}
\chi_{ij}^{(2)}(\bfq) = -\Pi_L(0) v_F^2 q_i q_j -\Pi_T(0) v_F^2(q^2 \delta_{ij} - q_i q_j),
\end{align}
with the coefficients
\begin{align}
\Pi_L(0) &= \frac{1}{\beta V} \sum_{\omega_n,\bfk} \left[ \frac{1}{4} D_0^2 +\frac{2}{3} D_0^3 (v_F k)^2 + \frac{2}{5} D_0^4 (v_F k)^4 \right] \label{eq:PI-L} \\
\Pi_T(0) &= \frac{1}{\beta V} \sum_{\omega_n,\bfk} \left[ \frac{3}{4} D_0^2 +\frac{2}{3} D_0^3 (v_F k)^2 + \frac{2}{15} D_0^4 (v_F k)^4 \right] . \label{eq:PI-T}
\end{align}

Let us evaluate the sum over the Matsubara frequency and the momentum to estimate the coefficients $\Pi_L$ and $\Pi_T$.
All we need to evaluate are three types of integrals appearing in Eqs.(\ref{eq:PI-L})(\ref{eq:PI-T}).
By introducing a function with a dimensionless parameter $\xi$,
\begin{align}
I(\xi) \equiv \frac{1}{\beta V} \sum_{\omega_n,\bfk} \frac{1}{\left[(i\omega_n+E_F)^2 - \xi v_F^2 k^2 \right]^2}
\end{align}
those three types of integrals can be expressed as
\begin{align}
\frac{1}{\beta V} \sum_{\omega_n,\bfk} D_0^2(i\omega_n,\bfk) &= I(1) \\
\frac{1}{\beta V} \sum_{\omega_n,\bfk} D_0^3(i\omega_n,\bfk) (v_F k)^2 &= \frac{1}{2} \frac{\partial I(\xi)}{\partial \xi} \Bigr|_{\xi=1} \\
\frac{1}{\beta V} \sum_{\omega_n,\bfk} D_0^4(i\omega_n,\bfk) (v_F k)^4 &= \frac{1}{6} \frac{\partial^2 I(\xi)}{\partial \xi^2} \Bigr|_{\xi=1} \ .
\end{align}
We can perform the Matsubara sum and the momentum integration in $I(\xi)$, yielding
\begin{align}
I(\xi) &= \frac{1}{\beta V} \sum_{\omega_n,\bfk} \frac{1}{\left[(i\omega_n+E_F)^2 - \xi v_F^2 k^2 \right]^2} \\
 &= \frac{1}{(2\pi)^3 \beta} \sum_{\omega_n} \int d^3\bfk \frac{1}{\left[(i\omega_n+E_F)^2 - \xi v_F^2 k^2 \right]^2} \\
 &= \frac{4\pi}{(2\pi)^3\beta} \sum_{\omega_n} \int_0^{k_C} dk \frac{k^2}{\left[(i\omega_n+E_F)^2 - \xi v_F^2 k^2 \right]^2} \\
 &= \frac{4\pi}{(2\pi)^3\beta v_F^2} \frac{\partial}{\partial \xi} \sum_{\omega_n} \int_0^{k_C} dk \frac{1}{(i\omega_n+E_F)^2 - \xi v_F^2 k^2} \\
 &= \frac{4\pi}{(2\pi)^3\beta v_F^2} \frac{\partial}{\partial \xi} \sum_{\omega_n} \int_0^{k_C} dk \frac{1}{2\sqrt{\xi}v_F k} \left[ \frac{1}{i\omega_n+E_F - \sqrt{\xi}v_F k} - \frac{1}{i\omega_n+E_F + \sqrt{\xi}v_F k} \right]\\
 &= \frac{4\pi}{(2\pi)^3 v_F^2} \frac{\partial}{\partial \xi} \int_0^{k_C} dk \frac{1}{2\sqrt{\xi}v_F k} \left[ f(\sqrt{\xi}v_F k -E_F) - f(-\sqrt{\xi}v_F k -E_F) \right]\\
 &= \frac{4\pi}{(2\pi)^3 v_F^2} \frac{\partial}{\partial \xi} \int_0^{k_C} dk \frac{-\theta(\sqrt{\xi}v_F k -E_F)}{2\sqrt{\xi}v_F k} \quad (\text{for } T=0)\\
 &= \frac{\pi}{(2\pi)^3 v_F^3} \int_0^{k_C} dk \left[ \frac{1}{\xi^{3/2} k} \theta(\sqrt{\xi}v_F k -E_F) - \frac{v_F}{\xi}\delta(\sqrt{\xi}v_F k -E_F) \right] \label{eq:I1} \\
 &= \frac{\pi}{(2\pi)^3 v_F^3} \frac{1}{\xi^{3/2}} \left[ \ln \frac{\sqrt{\xi} v_F k_C}{E_F} - 1 \right] 
 = \frac{1}{8\pi^2 v_F^3} \frac{1}{\xi^{3/2}} \left[ \ln \frac{\sqrt{\xi} k_C}{k_F} - 1 \right]
\label{eq:I2}
\end{align}
Here, the first (logarithmic) term in Eq.(\ref{eq:I1}) or Eq.(\ref{eq:I2}) comes from both valence and conduction bands,
corresponding to the Van Vleck contribution,
while the second (constant) term picks up the Fermi surface, corresponding to the Pauli contribution.
Thus we obtain the integrals
\begin{align}
& \frac{1}{\beta V} \sum_{\omega_n,\bfk} D_0^2 = \frac{1}{8\pi^2 v_F^3} \left[ \ln \frac{k_C}{k_F} - 1 \right], \quad 
\frac{1}{\beta V} \sum_{\omega_n,\bfk} D_0^3(v_F k)^2 = -\frac{1}{8\pi^2 v_F^3} \left[ \frac{3}{4}\ln \frac{k_C}{k_F} - 1 \right], \\
& \frac{1}{\beta V} \sum_{\omega_n,\bfk} D_0^4(v_F k)^4 = \frac{1}{8\pi^2 v_F^3} \left[ \frac{5}{8}\ln \frac{ K}{k_F} - \frac{23}{24} \right],
\end{align}
leading to the longitudinal and transverse coefficients
\begin{align}
\Pi_L(0) = \frac{1}{8\pi^2 v_F^3}\frac{1}{30}, \quad \Pi_T(0) = \frac{1}{8\pi^2 v_F^3}\left[\frac{1}{3}\ln\frac{k_C}{k_F} - \frac{19}{90}\right].
\end{align}
We can rearrange them into the ``isotropic'' and ``anisotropic'' exchange coefficients,
\begin{align}
J_\mathrm{i} =\frac{1}{2}\jsd^2 \Pi_L(0) = \frac{\jsd^2}{48\pi^2 v_F^3}\frac{1}{10}, \quad J_\mathrm{a} = \frac{1}{2}\jsd^2 [\Pi_T(0)-\Pi_L(0)] = \frac{\jsd^2}{48\pi^2 v_F^3 }\left[\ln\frac{k_C}{k_F} - \frac{11}{15}\right].
\end{align}

\subsection{Logarithmic divergence at $E_F=0$}
If the Fermi level is at the Dirac point $(E_F=0)$,
we can no longer rely on the power series expansion by $\bfq$ to evaluate the susceptibility tensor $\chi_{ij}(\bfq)$.
Here we have
\begin{align}
\chi_{ij}(\bfq) &= \frac{1}{4\beta V} \sum_{\lambda,\omega_n,\bfk} \Tr \left[ \sigma_i G_{0\lambda}(i\omega_n,\bfk+\tfrac{1}{2}\bfq) \sigma_j G_{0\lambda}(i\omega_n,\bfk-\tfrac{1}{2}\bfq) \right] \\
 &= \frac{1}{\beta V} \sum_{\omega_n,\bfk} \frac{(i\omega_n)^2 \delta_{ij} + v_F^2[2(k_i k_j -\tfrac{1}{4}q_i q_j) - (k^2-\tfrac{1}{4}q^2)\delta_{ij}] }{[(i\omega_n)^2 - v_F^2|\bfk+\tfrac{1}{2}\bfq|^2][(i\omega_n)^2 - v_F^2|\bfk-\tfrac{1}{2}\bfq|^2]} \\
 &= \frac{1}{\beta V} \sum_{\omega_n,\bfk} \frac{(i\omega_n)^2 \delta_{ij} + v_F^2[2(k_i k_j -\tfrac{1}{4}q_i q_j) - (k^2-\tfrac{1}{4}q^2)\delta_{ij}]}{ 2v_F^2 \bfk\cdot\bfq} \left[ \frac{1}{(i\omega_n)^2 - v_F^2|\bfk+\tfrac{1}{2}\bfq|^2} - \frac{1}{(i\omega_n)^2 - v_F^2|\bfk-\tfrac{1}{2}\bfq|^2} \right] \\
 &= \frac{1}{\beta V} \sum_{\omega_n,\bfk} \frac{1}{2v_F^2 \bfk\cdot\bfq} \label{eq:fraction} \\
 & \quad \quad \times \left[ \frac{v_F^2[(|\bfk+\tfrac{1}{2}\bfq|^2 - k^2+\tfrac{1}{4}q^2)\delta_{ij} + 2(k_i k_j -\tfrac{1}{4}q_i q_j)]}{(i\omega_n)^2 - v_F^2|\bfk+\tfrac{1}{2}\bfq|^2} - \frac{v_F^2[(|\bfk-\tfrac{1}{2}\bfq|^2 - k^2+\tfrac{1}{4}q^2)\delta_{ij} + 2(k_i k_j -\tfrac{1}{4}q_i q_j)]}{(i\omega_n)^2 - v_F^2|\bfk-\tfrac{1}{2}\bfq|^2} \right] \nonumber \\
 &= \frac{1}{\beta V} \sum_{\omega_n,\bfk} \frac{1}{\bfk\cdot\bfq} \frac{(|\bfk+\tfrac{1}{2}\bfq|^2 - k^2+\tfrac{1}{4}q^2)\delta_{ij} + 2(k_i k_j -\tfrac{1}{4}q_i q_j)}{(i\omega_n)^2 - v_F^2|\bfk+\tfrac{1}{2}\bfq|^2},
\end{align}
where we have changed the variable $\bfk \rightarrow -\bfk$ for the second term in Eq.(\ref{eq:fraction}).
By performing the Matsubara sum, we obtain
\begin{align}
\chi_{ij}(\bfq) &= \frac{-1}{V} \sum_{\bfk} \frac{1}{\bfk\cdot\bfq} \frac{(|\bfk+\tfrac{1}{2}\bfq|^2 - k^2+\tfrac{1}{4}q^2)\delta_{ij} + 2(k_i k_j -\tfrac{1}{4}q_i q_j)}{2v_F |\bfk+\tfrac{1}{2}\bfq|} \\
 &= -\frac{1}{(2\pi)^3} \int d^3\bfk \frac{1}{\bfk\cdot\bfq} \frac{(\bfk\cdot\bfq + \tfrac{1}{2}q^2)\delta_{ij} + 2(k_i k_j -\tfrac{1}{4}q_i q_j)}{2v_F |\bfk+\tfrac{1}{2}\bfq|}.
\end{align}

In order to complete the momentum integral,
we fix the external momentum $\bfq$ to the $z$-direction,
without losing the generality.
The integral then becomes
\begin{align}
\chi_{ij}(\bfq) &= -\frac{1}{(2\pi)^3} \int_{-k_C}^{k_C} d k_z \frac{1}{k_z q} \int_0^{\sqrt{k_C^2 -k_z^2}} d k_\perp \ 2\pi k_\perp \frac{(k_z q + \tfrac{1}{2}q^2)\delta_{ij} +2[(k_z^2 -\tfrac{1}{4}q^2)\delta_{ij}\delta_{iz} + \tfrac{1}{2} k_\perp^2 \delta_{ij}(\delta_{ix}+\delta_{iy}) ]}{2v_F\sqrt{k_\perp^2 + (k_z+\tfrac{1}{2}q)^2}} \\
&= -\frac{1}{8\pi^2} \int_{-k_C}^{k_C} d k_z \frac{1}{k_z q} \int _0^{k_C^2 -k_z^2} d\xi \ \frac{(2k_z^2+k_z q)\delta_{ij}\delta_{iz} + (k_z q +\tfrac{1}{2}q^2 +\xi)\delta_{ij}(\delta_{ix}+\delta_{iy})}{2v_F\sqrt{\xi + (k_z+\tfrac{1}{2}q)^2}} \quad (\delta_{ix}+\delta_{iy}+\delta_{iz}=1) \\
&= -\frac{1}{8\pi^2} \int_{-k_C}^{k_C} d k_z \frac{1}{k_z q} \int _0^{k_C^2 -k_z^2} d\xi \ \frac{2k_z(k_z+\tfrac{1}{2}q)\delta_{ij}\delta_{iz} + [\xi+(k_z+\tfrac{1}{2}q)^2 -k_z^2+\tfrac{1}{4}q^2]\delta_{ij}(\delta_{ix}+\delta_{iy})}{2v_F\sqrt{\xi + (k_z+\tfrac{1}{2}q)^2}} \\
&= -\frac{1}{8\pi^2} \int_{-k_C}^{k_C} d k_z \frac{2k_z(k_z+\tfrac{1}{2}q)\delta_{ij}\delta_{iz} -(k_z^2-\tfrac{1}{4}q^2)\delta_{ij}(\delta_{ix}+\delta_{iy}) }{v_F k_z q} \left[\sqrt{k_C^2-k_z^2 +(k_z+\tfrac{1}{2}q)^2} - |k_z+\tfrac{1}{2}q|\right] \nonumber \\
 & \quad \quad -\frac{1}{8\pi^2} \int_{-k_C}^{k_C} d k_z \frac{\delta_{ij}(\delta_{ix}+\delta_{iy})}{3 v_F k_z q}\left[  [k_C^2-k_z^2 +(k_z+\tfrac{1}{2}q)^2]^{3/2} - |k_z+\tfrac{1}{2}q|^3 \right].
\end{align}
Since $k_C^2 \gg k_z q, q^2$,
we can apply the power series expansion to $\sqrt{k_C^2-k_z^2 +(k_z+\tfrac{1}{2}q)^2} = \sqrt{k_C^2 +k_z q+\tfrac{1}{4}q^2}$,
which leads to
\begin{align}
\sqrt{k_C^2 +k_z q+\tfrac{1}{4}q^2} &= k_C + \frac{k_z q+\tfrac{1}{4}q^2}{2k_C} +\cdots \\
[k_C^2 +k_z q+\tfrac{1}{4}q^2]^{3/2} &= k_C^3 + \frac{3}{2} k_C (k_z q+\tfrac{1}{4}q^2) +\cdots.
\end{align}

We can see that the logarithmic divergence comes from the $k_z$-integral over the integrands of $O(1/k_z)$
around $k_z=0$.
If the integrand is an even function of $k_z$, it does not lead to the logarithmic divergence.
On the other hand, if the integrand is an odd function, it vanishes due to the symmetry of the integral.
Therefore the logarithmic divergence comes from the part of the integrand that cannot be classified as either even or odd,
which corresponds to the non-analytic terms with $|k_z+\tfrac{1}{2}q|$.
By extracting such terms of $O(1/k_z)$, we obtain
\begin{align}
\chi_{ij}(\bfq) &\approx \frac{1}{8\pi^2} \int_{-k_C}^{k_C} d k_z \frac{\tfrac{1}{4}q^2\delta_{ij}(\delta_{ix}+\delta_{iy}) }{v_F k_z q}|k_z+\tfrac{1}{2}q| + \frac{1}{8\pi^2} \int_{-k_C}^{k_C} d k_z \frac{\delta_{ij}(\delta_{ix}+\delta_{iy})}{3 v_F k_z q}|k_z+\tfrac{1}{2}q|^3 \\
 &= \frac{\delta_{ij}(\delta_{ix}+\delta_{iy})}{8\pi^2 v_F q} \int_{-k_C}^{k_C} d k_z \left[\frac{q^2}{4} \frac{|k_z+\tfrac{1}{2}q|}{k_z} + \frac{1}{3} \frac{|k_z+\tfrac{1}{2}q|^3}{k_z}\right] \\
 &= \frac{\delta_{ij}(\delta_{ix}+\delta_{iy})}{8\pi^2 v_F q} \left\{ \int_{-q/2}^{k_C} d k_z \left[\frac{q^2}{4} \frac{k_z+\tfrac{1}{2}q}{k_z} + \frac{1}{3} \frac{(k_z+\tfrac{1}{2}q)^3}{k_z}\right] + \int_{-k_C}^{-q/2} d k_z \left[\frac{q^2}{4} \frac{-(k_z+\tfrac{1}{2}q)}{k_z} + \frac{1}{3} \frac{-(k_z+\tfrac{1}{2}q)^3}{k_z}\right] \right\} \\
 &\approx \frac{\delta_{ij}(\delta_{ix}+\delta_{iy})}{8\pi^2 v_F q} \left\{  \int_{-q/2}^{k_C} d k_z \frac{q^3}{6} \frac{1}{k_z} - \int_{-k_C}^{-q/2} d k_z \frac{q^3}{6} \frac{1}{k_z} \right\} \\
 &= \delta_{ij}(\delta_{ix}+\delta_{iy}) \frac{q^2}{24\pi^2 v_F} \ln \left|\frac{k_C}{q/2}\right| \\
 &\approx \delta_{ij}(\delta_{ix}+\delta_{iy}) \frac{1}{48\pi^2 v_F} q^2\ln q^{-2}.
\end{align}

Thus the gradient part of the free energy $F_1[\bfM]$ becomes
\begin{align}
F_1[\bfM] &= \sum_{\bfq} \sum_{i,j=x,y,z} \frac{J_\mathrm{sd}^2}{2} \chi_{ij}(\bfq) M_i(\bfq) M_j(-\bfq) \\
 &= \sum_{\bfq} \left\{ \frac{J_\mathrm{sd}^2}{96\pi^2 v_F} \ln q^{-2} \left[q^2 \bfM(\bfq)\cdot\bfM(-\bfq) - (\bfq\cdot\bfM(\bfq))(\bfq\cdot\bfM(-\bfq))\right] + O(q^2) \right\} \\
 &= \sum_{\bfq} \left\{ \frac{J_\mathrm{sd}^2}{96\pi^2 v_F} \ln q^{-2} [\bfq\times\bfM(\bfq)]\cdot[\bfq\times\bfM(-\bfq)] + O(q^2) \right\},
\end{align}
where the first term in $\{ \cdots \}$ corresponds to the anisotropic correlation term,
characterized by the coefficient $J_\mathrm{a}$ defined for $k_F \gg q$.
Therefore, we conclude that the anisotropic correlation term in the free energy is of the order $O(q^2 \ln q^{-1})$,
if the Fermi level is close to the Dirac points.

\section{Spin wave dispersion}
The spin waves in the ferromagnetic phase polarized in the $z$-direction,
namely the fluctuations of spins around the mean field, can well be described by the bosonic Holstein--Primakoff fields $(z^\dag,z)$, which are introduced by the transformation
\begin{align}
S_z(\bfr) &= M - z^\dag(\bfr) z(\bfr), \\
S_+(\bfr) &= \sqrt{2M-z^\dag(\bfr)z(\bfr)} \ z(\bfr) \simeq \sqrt{2M}z(\bfr), \label{eq:S+}\\
S_-(\bfr) &= z^\dag(\bfr) \sqrt{2M-z^\dag(\bfr)z(\bfr)} \simeq \sqrt{2M}z^\dag(\bfr) \label{eq:S-}
\end{align}
in the operator formalism.
Here the approximations in Eqs.(\ref{eq:S+})(\ref{eq:S-}) are justified for $M \gg 1$.

In the path integral formalism, the total action is given with the fermionic itinerant carrier field $(\psi^\dag,\psi)$ and the bosonic magnon field $(z^\dag,z)$ as
\begin{align}
S_\mathrm{tot} &= S_\mathrm{MF}[\psi^\dag,\psi] + S_\mathrm{mag}[z^\dag,z] + S_I[\psi^\dag,\psi;z^\dag,z] \\
S_\mathrm{MF}[\psi^\dag,\psi] &= \int_0^\beta d\tau \sum_{\lambda,\bfk} \psi_\lambda^\dag(\tau,\bfk) \left[\partial_\tau-E_F +  {H}_{0\lambda}(\bfk) -\frac{1}{2}\Delta {\sigma}_z \right] \psi_\lambda(\tau,\bfk) \\
 &= -\sum_{\lambda,\omega_n,\bfk} \psi_\lambda^\dag(i\omega_n,\bfk) \left[(i\omega_n-E_F) -  {\bm{\sigma}}\cdot \bm{g}_\lambda(\bfk) \right] \psi_\lambda(i\omega_n,\bfk) \\
 &\equiv -\sum_{\lambda,\omega_n,\bfk} \psi_\lambda^\dag(i\omega_n,\bfk) \left[ \bar{G}_\lambda(i\omega_n,\bfk)\right]^{-1} \psi_\lambda(i\omega_n,\bfk) \\
S_\mathrm{mag}[z^\dag,z] &= \int_0^\beta d\tau \int d^3\bfr \  z^\dag(\tau,\bfr) \partial_\tau z(\tau,\bfr) = \sum_{\nu_m,\bfq} z^\dag(i\nu_m,\bfq)[-i\nu_m]z(i\nu_m,\bfq) \\
S_I[\psi^\dag,\psi;z^\dag,z] &= \int_0^\beta d\tau \int d^3\bfr \  \frac{J_\mathrm{sd}}{2} \left[\sqrt{2M}(z {s}^-+z^\dag {s}^+) - 2z^\dag z  {s}_z\right], \quad (s^\pm = (\sigma_x\pm\sigma_y)/\sqrt{2}). \label{eq:interaction}
\end{align}
Since $H_M = (1/2\chi_M)\int d^3 \bfr |\bfM(\bfr)|^2$ depends only on the amplitude of magnetization $|\bfM|$,
it does not depend on $(z^\dag,z)$, namely the phase fluctuation of $\bfM(\bfr)$.
Here fermionic and bosonic Matsubara frequencies are given by
$\omega_n = (2\pi/\beta)(n+1/2)$ and $\nu_m = (2\pi/\beta)m$ respectively,
and
\begin{align}
\bm{g}_\lambda(\bfk) = (\lambda v_F k_x, \lambda v_F k_y, \lambda v_F k_z-\frac{1}{2}\Delta)
\end{align}
 is the spin decomposition of the mean-field Hamiltonian,
with $\Delta = J_\mathrm{sd} M$ the mean-field spin splitting energy.
Since the 
Mean-field propagator $ \bar{G}_\lambda$ can be decomposed as
\begin{align}
 \bar{G}_\lambda(i\omega_n,\bfk) = \bar{D}_\lambda(i\omega_n,\bfk)\left[(i\omega_n-E_F)+ {\bm{\sigma}}\cdot \bm{g}_\lambda(\bfk)\right],
\end{align}
with the denominator
\begin{align}
\bar{D}_\lambda(i\omega_n,\bfk) = \left[(i\omega_n-E_F)^2 -E_\lambda^2(\bfk)\right]^{-1}, \quad E_\lambda(\bfk) = |\bm{g}_\lambda(\bfk)|.
\end{align}
The interaction term $S_I$ consists of three-point vertices $z \psi^\dag \psi$ and $z^\dag \psi^\dag \psi$ for in-plane spin components and four-point vertex $z^\dag z \psi^\dag \psi$ for out-of-plane components.

The magnon effective action $S_\mathrm{eff}[z^\dag,z]$ can be obtained by integrating out the fermionic fields,
\begin{align}
e^{-S_\mathrm{eff}[z^\dag,z]} = \int [d\psi^\dag d\psi] e^{-S_\mathrm{tot}[\psi^\dag,\psi;z^\dag,z]}.
\end{align}
Taking $S_I$ as a perturbation, the first- and second-order corrections to $S_\mathrm{eff}$ are given by
\begin{align}
S_\mathrm{eff}^{(1)}[z^\dag,z] &=  \sum_{\nu_m,\bfq} z^\dag(i\nu_m,\bfq) \Sigma_z z(i\nu_m,\bfq) \\
S_\mathrm{eff}^{(2)}[z^\dag,z] &= -\frac{\jsd^2}{2} \sum_{\nu_m,\bfq} Z^\dag(i\nu_m,\bfq) \bar{\chi}(i\nu_m,\bfq) Z(i\nu_m,\bfq),
\end{align}
where the ``spinor'' representation $Z$ of the magnon field is defined by
\begin{align}
Z(i\nu_m,\bfq) = \left(\begin{array}{c} z(i\nu_m,\bfq) \\ z^\dag(-i\nu_m,-\bfq) \end{array}\right).
\end{align}
The self-energy $\Sigma_z$ and the spin susceptibility tensor $\bar{\chi}$ are given by
\begin{align}
\Sigma_z &= -\frac{\jsd}{\beta V} \sum_{\lambda,\omega_n,\bfk} \Tr\left[ {s}_z  \bar{G}_\lambda(i\omega_n,\bfk)\right] \\
\bar{\chi}(i\nu_m,\bfq) &= \left( \begin{array}{cc}
 \bar{\chi}_{+-}(i\nu_m,\bfq) & \bar{\chi}_{++}(i\nu_m,\bfq) \\
 \bar{\chi}_{--}(i\nu_m,\bfq) & \bar{\chi}_{-+}(i\nu_m,\bfq)
\end{array} \right) \\
\bar{\chi}_{\alpha\beta}(i\nu_m,\bfq) &= -\frac{ M}{2\beta V} \sum_{\lambda,\omega_n,\bfk} \Tr\left[ {s}^\alpha  \bar{G}_\lambda(i\omega_n+i\nu_m,\bfk+\bfq)  {s}^\beta  \bar{G}_\lambda(i\omega_n,\bfk)\right]. \quad (\alpha,\beta = +,-)
\end{align}
Using this form, the magnon effective action up to one-loop of the Dirac electron is written as
\begin{align}
S_\mathrm{eff}[z^\dag,z] &= \frac{1}{2} \sum_{\nu_m,\bfq} Z^\dag(i\nu_m,\bfq) {D}_M^{-1}(i\nu_m,\bfq) Z(i\nu_m,\bfq) \\
{D}_M^{-1}(i\nu_m,\bfq) &= \left( \begin{array}{cc}
 - i\nu_m + \Sigma_z - \jsd^2 \bar{\chi}_{+-}(i\nu_m,\bfq) & -\jsd^2 \bar{\chi}_{++}(i\nu_m,\bfq) \\
 -\jsd^2 \bar{\chi}_{--}(i\nu_m,\bfq) & i\nu_m + \Sigma_z - \jsd^2 \bar{\chi}_{-+}(i\nu_m,\bfq)
\end{array} \right)
\end{align}
The gap of the magnon is given by the pole of the magnon propagator at $\bfq=0$,
i.e. the solution of the equation $ {D}_M^{-1}(i\nu_m,0)=0$.
Since the susceptibility tensor $ {\chi}$ does not depend linearly on $i\nu_m$ and is diagonal at $\bfq=0$, as we shall see below,
the magnon gap $\Delta_M$ is given by a simple relation,
\begin{align}
\Delta_M = \Sigma_z - \jsd^2 \bar{\chi}_{+-}(i\nu_m=0,\bfq=0)
\end{align}
up to the fermion one-loop.

\subsection{Vanishing gap of magnon}

Let us first evaluate the magnon self-energy part, given by $\Sigma_z$.
It can be simplified by taking the trace and the Matsubara sum,
\begin{align}
\Sigma_z &= -\frac{\jsd}{\beta V} \sum_{\lambda,\omega_n,\bfk} \bar{D}_\lambda(i\omega_n,\bfk) g_z^\lambda(\bfk) \\
 &= -\frac{\jsd}{\beta V} \sum_{\lambda,\omega_n,\bfk} \frac{g_z^\lambda(\bfk)}{2E_\lambda(\bfk)}\left[\frac{1}{i\omega_n+E_F-E_\lambda(\bfk)} - \frac{1}{i\omega_n+E_F+E_\lambda(\bfk)}\right] \\
 &= -\frac{\jsd}{V} \sum_{\lambda,\bfk} \frac{g_z^\lambda(\bfk)}{2E_\lambda(\bfk)} \left[f(E_\lambda(\bfk)-E_F) - f(-E_\lambda(\bfk)-E_F)\right] \\
 &= -\frac{\jsd}{2}\mathcal{S}_z ,
\end{align}
with $\mathcal{S}_z$ the average spin polarization per one site,
\begin{align}
\mathcal{S}_z = \frac{1}{V}\sum_{\lambda,\bfk} \frac{g_z^\lambda(\bfk)}{E_\lambda(\bfk)} \left[f(E_\lambda(\bfk)-E_F) - f(-E_\lambda(\bfk)-E_F)\right].
\end{align}
If the cutoff $K$ is sufficiently large such that the Fermi surface is fully covered within the cutoff region $(E_F \ll v_F K)$,
the conduction band $f(E_\lambda(\bfk)-E_F)$ contribution vanishes because it is symmetric around the shifted Dirac point $k_z=-\lambda \Delta/v_F$.
Therefore, we only need to consider the Fermi sea contribution regardless of the Fermi level $E_F$ as long as it is above the Dirac point,
\begin{align}
\Sigma_z &= \frac{\jsd}{2} \sum_\lambda \frac{1}{(2\pi)^3} \int_{|\bfk|<k_C} d^3\bfk \frac{g_z^\lambda(\bfk)}{E_\lambda(\bfk)} \\
 &= \frac{\jsd}{2} \frac{2}{v_F^3} \int_{|\bfk|<k_C} d^3\bfk \frac{v_F k_z +\frac{\Delta}{2}}{\sqrt{(v_F k_\perp)^2 + (v_F k_z +\frac{\Delta}{2})^2}} \\
 &= \frac{\jsd}{2}\left(\delta-\frac{\delta^3}{20}\right),
\end{align}
where $\delta = \Delta/v_F k_C$.

Next we evaluate the susceptibility tensor, first in the static limit $i\nu_m=0, \ \bfq=0$,
which contributes to the gap of the spin wave.
From the above discussion, the only remaining part is the diagonal ($+-$ or $-+$) components,
given by
\begin{align}
\bar{\chi}_{+-}(0) &= -\frac{M}{2\beta V} \sum_{\lambda,\omega_n,\bfk} \Tr \left[ {s}_+  \bar{G}_\lambda(i\omega_n,\bfk)  {s}_-  \bar{G}_\lambda(i\omega_n,\bfk) \right] \\
 &= -\frac{M}{2\beta V} \sum_{\lambda,\omega_n,\bfk} \bar{D}_\lambda^2(i\omega_n,\bfk) \left[(i\omega_n+E_F)^2 - (g_z^\lambda(\bfk))^2\right] \\
 &= -\frac{M}{2\beta V} \sum_{\lambda,\omega_n,\bfk} \left[ \bar{D}_\lambda(i\omega_n,\bfk) + \bar{D}_\lambda^2(i\omega_n,\bfk)|g_\perp^\lambda(\bfk)|^2 \right]
\end{align}
Evaluating the Matsubara sum, we obtain
\begin{align}
\bar{\chi}_{+-}(0) &= - \frac{M}{2 V} \sum_{\lambda,\bfk} \left( \frac{f_{\lambda +} - f_{\lambda -}}{2E_\lambda(\bfk)} +\left[ \frac{f'_{\lambda +} + f'_{\lambda -}}{4 E_\lambda^2(\bfk)} - \frac{f_{\lambda +} - f_{\lambda -}}{4E_\lambda^3(\bfk)} \right]|g_\perp^\lambda(\bfk)|^2 \right),
\end{align}
where we use the shorthand notation $f_{\lambda\pm}=f(\pm E_\lambda(\bfk)-E_F)$.
$f'_{\lambda +}$ gives the Fermi surface contribution, which cancels with the Fermi sea contribution from the conduction band $(f_{\lambda +})$ due to the spherical symmetry around the shifted Dirac point, as in the calculation of $\Sigma_z$.
Therefore, all we need to evaluate is the valence band Fermi sea contribution, given by
\begin{align}
\bar{\chi}_{+-}(0) &= \frac{M}{2(2\pi)^3} \sum_{\lambda} \int_{|\bfk|<k_C} d^3\bfk \left( \frac{1}{2E_\lambda(\bfk)} - \frac{1}{4E_\lambda^3(\bfk)} |g_\perp^\lambda(\bfk)|^2 \right) = \frac{\jsd}{2}\left[\delta - \frac{\delta^3}{20}\right].
\end{align}

From the above calculations, we find that the magnon gap up to one-loop order of Dirac electron vanishes,
\begin{align}
\Delta_M = \Sigma_z - \jsd^2 \bar{\chi}_{+-}(i\nu_m=0,\bfq=0) = 0.
\end{align}

We can establish a Ward--Takahashi identity up to fermion one-loop that justifies the above relation.
Since we are interested in the magnon two-point Green's function with the external frequency and momentum zero,
we can start from the total action $S_\mathrm{tot}^{(0)}[\psi^\dag,\psi;z_0^\dag,z_0]$,
where we define $z_0^{(\dag)} = z^{(\dag)}(i\nu_m=0,\bfq=0)$ and set all the other modulating magnon fields to zero,
given by
\begin{align}
S_\mathrm{tot}^{(0)}[\psi^\dag,\psi;z_0^\dag,z_0] &= -\sum_{\lambda,\omega_n,\bfk} \psi_\lambda^\dag(i\omega_n,\bfk) \left[ i\omega_n +E_F -  {\bm{\sigma}} \cdot \left(\lambda v_F \bfk - \frac{\jsd}{2} \bfS(z_0^\dag,z_0)\right) \right] \psi_\lambda(i\omega_n,\bfk).
\end{align}
Here $\bfS(z_0^\dag,z_0)$ is the spin field composed of the Holstein--Primakoff field $(z_0^\dag,z_0)$,
which is considered as a mean-field-like $c$-number at this moment.
The mean-field effective action (free energy) is given by integrating out the fermion fields,
\begin{align}
\Gamma^{(0)}(z_0^\dag,z_0) &= -\frac{1}{\beta V} \int [d\psi^\dag d\psi] e^{-S_\mathrm{tot}^{(0)}[\psi^\dag,\psi;z_0^\dag,z_0]} \\
 &= -\frac{1}{\beta V} \sum_{\lambda,\omega_n,\bfk} \ln \det \left[ i\omega_n +E_F -  {\bm{\sigma}} \cdot \left(\lambda v_F \bfk - \frac{\jsd}{2} \bfS(z_0^\dag,z_0)\right) \right], \label{eq:mf-effective-action}
\end{align}
and its second derivative yields the inverse of the magnon propagator at $i\nu_m =\bfq=0$,
\begin{align}
\left[D_M(0)\right]^{-1} = \Delta_M = -\frac{\partial^2 \Gamma^{(0)}(z_0^\dag,z_0)}{\partial z_0^\dag \partial z_0} \Biggr|_{z_0^\dag=z_0=0} .
\end{align}
The determinant in Eq.(\ref{eq:mf-effective-action}) reads
\begin{align}
(i\omega_n+E_F)^2 - \left| \lambda v_F \bfk - \frac{\jsd}{2} \bfS(z_0^\dag,z_0) \right|^2 = (i\omega_n+E_F)^2 -\left[ (v_F k)^2 + \frac{\jsd^2}{4} |\bfS(z_0^\dag,z_0)|^2 - \lambda \jsd v_F k_S |\bfS(z_0^\dag,z_0)| \right],
\end{align}
where $k_S$ is the wave vector component projected on $\bfS(z_0^\dag,z_0)$.
Therefore, if the $\bfk$-integral is spherically symmetric,
the effective action $\Gamma^{(0)}$ depends only on the magnitude of the spin, $|\bfS(z_0^\dag,z_0)|^2$,
not on its direction.
If we use the approximated Holstein--Primakoff field defined in Eq.(\ref{eq:interaction}),
the spin magnitude reads
\begin{align}
|\bfS(z_0^\dag,z_0)|^2 = S_z^2 + S_+ S_- = (M-z_0^\dag z_0)^2 + 2M z_0^\dag z_0 = M^2 +(z_0^\dag z_0)^2,
\end{align}
which is constant up to $O(z_0^\dag z_0)$,
i.e. justified for $M \gg z_0^\dag z_0 \sim M-M_z$.
Therefore, the second derivative $\partial^2 \Gamma^{(0)}\left(|\bfS(z_0^\dag,z_0)|^2\right)/\partial z_0^\dag \partial z_0$ vanishes at the lowest order, yielding the gapless spin-wave mode.
Since our Holstein--Primakoff field breaks the spin symmetry at higher orders,
this argument cannot be applied to the magnon vertex functions more than three points.
It cannot either be applied to the processes with internal magnons carrying finite frequency or momentum.

\subsection{Magnon dispersion}
We now investigate the dispersion relation of the spin waves, by evaluating the susceptibility tensor $\bar{\chi}_{\alpha\beta}(\bfq)$.
This tensor is constructed from the single-particle Dirac propagator $\bar{G}$ in the presence of the mean field $\Delta$.
Taking the exchange splitting $\Delta$
the mean-field propagator $\bar{G}$ can simply be replaced by the free propagator $G_0$.
Thus we can reconstruct the tensors $\bar{\chi}_{\alpha\beta}$ from the tensors $\chi_{ij}$ obtained in deriving the effective field theory,
as
\begin{align}
\bar{\chi}_{\alpha\beta}(\bfq) &= \frac{M}{2} \left[\chi_{xx}(\bfq) + i\alpha \chi_{yx}(\bfq) + i\beta \chi_{xy}(\bfq) - \alpha\beta \chi_{yy}(\bfq) \right] + O(J_\mathrm{sd}^3) \quad (\alpha,\beta = \pm)
\end{align}
Let us again limit ourselves to the case $|\bfq|\ll k_F$
and rely on the power series expansion by $\bfq$,
which yields
\begin{align}
\chi_{ij}(\bfq) &= \chi_{ij}(0) - [\Pi_L(0)-\Pi_T(0)] q_i q_j - \Pi_T(0) q^2\delta_{ij} +O(q^4) \\
\bar{\chi}_{\pm \mp}(\bfq) &= \bar{\chi}_{\pm \mp}(0) - \frac{M}{2} [\Pi_L(0)-\Pi_T(0)] q_\perp^2 - M \Pi_T(0) q^2 +O(q^4) \\
 &= \bar{\chi}_{\pm \mp}(0) - \frac{M}{2} [\Pi_L(0)+\Pi_T(0)] q_\perp^2 - M \Pi_T(0) q_z^2 +O(q^4)\\
\bar{\chi}_{\pm \pm}(\bfq) &= \bar{\chi}_{\pm \pm}(0) - \frac{M}{2} [\Pi_L(0)-\Pi_T(0)] (q_\pm)^2+O(q^4),
\end{align}
where $q_\pm = q_x \pm i q_y$ and $q_\perp = \sqrt{q_x^2+q_y^2} = |q_\pm|$.
Thus the effective propagator matrix of magnon reads
\begin{align}
& D_M(i\nu_m,\bfq) = \\
& \left(
\begin{array}{cc}
-i\nu_m +\frac{1}{2} \jsd^2 M[\Pi_L(0)+\Pi_T(0)] q_\perp^2 + \jsd^2 M \Pi_T(0) q_z^2 & \frac{1}{2} \jsd^2  M[\Pi_L(0)-\Pi_T(0)] (q_-)^2  \\
\frac{1}{2}\jsd^2  M[\Pi_L(0)-\Pi_T(0)] (q_+)^2 & i\nu_m +\frac{1}{2}\jsd^2  M[\Pi_L(0)+\Pi_T(0)] q_\perp^2 +\jsd^2  M \Pi_T(0) q_z^2
\end{array}
\right)^{-1} +O(q^4) \nonumber
\end{align}
The magnon dispersion is given by the pole of the propagator,
with the Wick rotation to the real-time formalism as $i\nu_m \rightarrow \epsilon$:
\begin{align}
\epsilon(\bfq) &= \sqrt{\left[\frac{\jsd^2  M}{2} [\Pi_L(0)+\Pi_T(0)] q_\perp^2 + \jsd^2  M \Pi_T(0) q_z^2\right]^2 - \left|\frac{\jsd^2 M}{2} [\Pi_L(0)-\Pi_T(0)] (q_+)^2\right|^2}+O(q^4) \\
 &= \sqrt{(\mu_\perp q_\perp^2 + \mu_z q_z^2)^2 - (\mu'_\perp q_\perp^2)^2}+O(q^4).
\end{align}
Here the coefficients are evaluated as 
\begin{align}
\mu_z & = \jsd^2 M \Pi_T(0) = \frac{\jsd \Delta}{8\pi^2 v_F^3} \left[\frac{1}{3}\ln\frac{k_C}{k_F} - \frac{19}{90}\right] \\
\mu_\perp &= \frac{\jsd^2 M}{2}[\Pi_T(0)+\Pi_L(0)] = \frac{\jsd \Delta}{8\pi^2 v_F^3} \left[\frac{1}{6}\ln\frac{k_C}{k_F} - \frac{4}{45}\right] \\
\mu'_\perp &= \frac{\jsd^2 M}{2}[\Pi_T(0)-\Pi_L(0)] = \frac{\jsd \Delta}{8\pi^2 v_F^3} \left[\frac{1}{6}\ln\frac{k_C}{k_F} - \frac{11}{90}\right],
\end{align}
where we have used the relation $\Delta = \jsd M$.

\end{widetext}

\begin{thebibliography}{99}
\bibitem{Cutoff}
Since the summation over $\bfk$ is UV-divergent,
it is quantitatively sensitive to the shift of the variable $\bfk$,
namely the shift of the cutoff.
In realistic systems, it is uniquely regularized by the shape of the Brillouin zone.
Here we take a spherically symmetric cutoff to meet the spin-orbital rotational symmetry.
\end{thebibliography}
\end{document}